\documentclass[final,5p,times,twocolumn]{elsarticle}

\usepackage{epsfig}
\usepackage{multirow}
\usepackage{tabularx}
\usepackage{amssymb}

\journal{Physics Letters B}

\begin{document}

\begin{frontmatter}

\title{The $^{25}$Mg(p,$\gamma$)$^{26}$Al reaction at low astrophysical energies}

\author[RUB]{F.~Strieder}\corref{corauthor}
\author[Napoli]{B.~Limata}
\author[LNGS]{A.~Formicola}
\author[Napoli]{G.~Imbriani}
\author[LNGS]{M.~Junker}
\author[Dresden]{D.~Bemmerer}
\author[RUB]{A.~Best\fnref{ND}}
\author[INFNPa]{C.~Broggini}
\author[INFNPa]{A.~Caciolli}
\author[INFNGe]{P.~Corvisiero}
\author[INFNGe]{H.~Costantini}
\author[Napoli]{A.~DiLeva}
\author[ATOMKI]{Z.~Elekes}
\author[ATOMKI]{Zs.~F\"ul\"op}
\author[Torino]{G.~Gervino}
\author[Milano]{A.~Guglielmetti}
\author[LNGS]{C.~Gustavino}
\author[ATOMKI]{Gy.~Gy\"urky}
\author[INFNGe]{A.~Lemut\fnref{LBNL}}
\author[Dresden]{M.~Marta\fnref{GSI}}
\author[Milano]{C.~Mazzocchi\fnref{Poland}}
\author[INFNPa]{R.~Menegazzo}
\author[INFNGe]{P.~Prati}
\author[Napoli]{V.~Roca}
\author[RUB]{C.~Rolfs}
\author[INFNPa]{C.~Rossi Alvarez}
\author[ATOMKI]{E.~Somorjai}
\author[Teramo]{O.~Straniero}
\author[NapoliII]{F.~Terrasi}
\author[RUB]{H.P.~Trautvetter}

\address[RUB]{Institut f\"ur Experimentalphysik, Ruhr-Universit\"at Bochum, D-44780 Bochum, Germany}
\address[Napoli]{Dipartimento di Scienze Fisiche, Universit\`a di Napoli ''Federico II'', and INFN Sezione di Napoli, I-80126 Napoli, Italy}
\address[LNGS]{INFN, Laboratori Nazionali del Gran Sasso (LNGS), I-67010 Assergi, Italy}
\address[Dresden]{Helmholtz-Zentrum Dresden-Rossendorf, Bautzner Landstr. 400, D-01328 Dresden, Germany}
\address[INFNPa]{Istituto Nazionale di Fisica Nucleare (INFN) Sezione di Padova, via Marzolo 8, I-35131 Padova, Italy}
\address[INFNGe]{Universit\`a di Genova and INFN Sezione di Genova, I-16146 Genova, Italy}
\address[ATOMKI]{Institute of Nuclear Research (ATOMKI), H-4026 Debrecen, Hungary}
\address[Torino]{Dipartimento di Fisica Sperimentale, Universit\`a di Torino and INFN Sezione di Torino, I-10125 Torino, Italy}
\address[Milano]{Universit\`a degli Studi di Milano and INFN Sezione di Milano, I-20133 Milano, Italy}
\address[Teramo]{Osservatorio Astronomico di Collurania, I-64100 Teramo, Italy}
\address[NapoliII]{Seconda Universit\`a di Napoli, I-81100 Caserta, and INFN Sezione di Napoli, I-80126 Napoli, Italy}


\cortext[corauthor]{Corresponding author: strieder@ep3.rub.de}
\fntext[ND]{present address: Deparment of Physics, University of Notre Dame, Notre Dame, IN 46556, USA}
\fntext[LBNL]{present address: Nuclear Science Division, Lawrence Berkeley National Laboratory, Berkeley, CA 94720, USA}
\fntext[GSI]{present address: GSI Helmholtzzentrum f\"ur Schwerionenforschung GmbH, D-64291 Darmstadt, Germany}
\fntext[Poland]{present address: Inst. of Experimental, Physics University of Warsaw, ul. Hoza 69, P-00-682 Warszawa, Poland}

\begin{abstract}

In the present work we report on a new measurement of resonance strengths $\omega\gamma$ in the reaction $^{25}$Mg(p,$\gamma$)$^{26}$Al at ${\rm E_{cm}}= 92$ and 189~keV. This study was performed at the LUNA facility in the Gran Sasso underground laboratory using a 4$\pi$ BGO summing crystal. For the first time the 92~keV resonance was directly observed and a resonance strength $\omega\gamma=(2.9\pm0.6)\times10^{-10}$~eV was determined. Additionally, the $\gamma$-ray branchings and strength of the 189~keV resonance were studied with a high resolution HPGe detector yielding an $\omega\gamma$ value in agreement with the BGO measurement, but 20~\% larger compared to previous works.

\end{abstract}

\begin{keyword}

Nuclear Astrophysics \sep Mg-Al cycle \sep proton-capture reaction \sep reaction rate

\end{keyword}

\end{frontmatter}

\section{Introduction}

The Mg-Al cycle plays a relevant role in the synthesis of Mg and Al isotopes and is activated in the H-burning regions of stars, when the temperature exceeds ${\rm T}\approx30 - 40 \times10^6$~K. This condition takes place in H-burning convective cores of the most massive main-sequence stars (T up to $50\times10^6$~K) \cite{Palacios05_AA,Limongi06_ApJ},  in H-burning shells of off-main-sequence stars of any mass (T up to 10$^8$~K) \cite{Cavallo98_ApJ,Izzard07_AA} and in explosive H-burning during Nova-like outbursts (T up to $4\times10^8$~K) \cite{Iliadis02_ApJS}.

An understanding of these nucleosynthesis processes is important for many astronomical observations: (a) the Mg-Al (anti)correlation observed in Globular Cluster stars \cite{Cavallo98_ApJ,Carretta10_AA}; (b) the well-known 1.809 MeV $\gamma$-ray line commonly associated with the radioactive decay of $^{26}$Al produced by the present generation of galactic massive stars \cite{diehl95,knodlesser99,Diehl06_Natur}; (c) the fossil records of $^{26}$Al derived from the isotopic composition of Mg observed in different phases of Ca-Al rich inclusions of carbonaceous chondrites \cite{Lee77_ApJ,Wasserburg85,Gallino04_NewAR} and in presolar grains \cite{Hoppe94_ApJ}.

In general, the Mg-Al cycle is initiated by the $^{24}$Mg(p,$\gamma$)$^{25}$Al reaction followed by a $\beta$-decay into $^{25}$Mg. Subsequently another proton capture creates $^{26}$Al: one of the most important radioactive isotopes for $\gamma$-ray astronomy.
The reaction $^{25}$Mg(p,$\gamma$)$^{26}$Al proceeds with about 60 - 80~\% probability, depending on the interaction energy, to the ground state of $^{26}$Al ($t_{1/2}\approx7\times10^5$ y) while the remaining 20 - 40~\% populate the 228 keV isomeric state ($t_{1/2}= 6$~s). The $^{26}$Al ground state decays into the first excited state of $^{26}$Mg giving rise to the 1.809 MeV $\gamma$-ray. Further proton captures on $^{26}$Al and $^{26}${Mg} form $^{27}$Al either directly or through a $^{27}${Si} $\beta$-decay. Finally, the $^{27}$Al(p,$\alpha)^{24}$Mg reaction closes the Mg-Al cycle.

The stellar evolution models rely on a firm evaluation of the reaction rates of the involved nuclear reactions, and in particular of $^{25}$Mg(p,$\gamma$)$^{26}$Al at the relevant temperatures for the activation of the Mg-Al cycle.

The reaction $^{25}$Mg(p,$\gamma$)$^{26}$Al ($\rm Q=6306$~keV) is dominated by isolated resonances although the high level density of $^{26}$Al seriously complicates the study of this reaction. In the past $^{25}$Mg(p,$\gamma$)$^{26}$Al resonances down to $\rm E=189$~keV (all energies are given in the center-of-mass frame except where quoted differently) have been studied in direct and indirect experiments \cite{Betts78_NPA,CH83a,CH83b,EN86,CH86,EN87,CH89,Rollefson90_NPA,IL90,IL96,PO98,arazi} while resonance strengths at lower energies have been deduced by indirect studies only \cite{CH89,Rollefson90_NPA,IL96}. These data suggest that resonances at $\rm E = 58$, 92, 189, and 304~keV play a significant role in the astrophysically relevant temperature range of H-burning in stars.
The 58~keV resonance is inaccessible for direct experiments, while there are additional resonances reported in literature, e.g. one at $\rm E = 130$~keV, which only contribute to the reaction rate in case the strengths are underestimated by 2 - 3 orders of magnitude.

Note the $^{25}$Mg(p,$\gamma)^{26}$Al reaction is also active in advanced burning phases, i.e. carbon and neon burning where for instance a major contribution to the galactic $^{26}$Al is produced \cite{Iliadis11_ApJSup}, but
these stages require much higher temperatures and are not effected by the results of the present study.

\begin{figure*}
\includegraphics[angle=0,width=\textwidth]{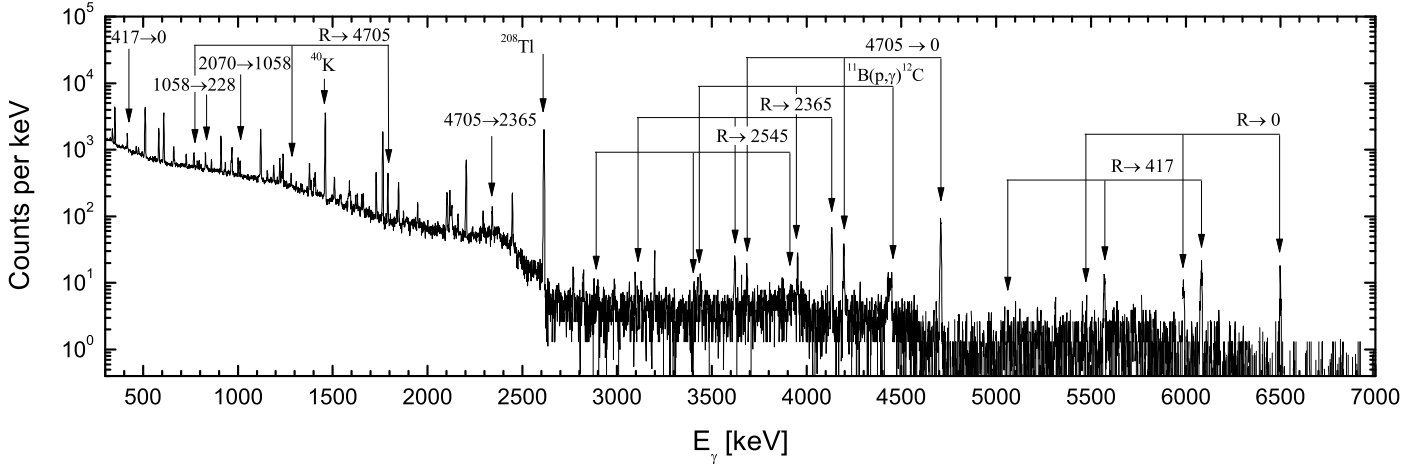}
\caption{The HPGe $\gamma$-ray spectrum taken at the ${\rm E}= 189$~keV $^{25}$Mg(p,$\gamma$)$^{26}$Al resonance showing the most prominent primary transitions, important secondary transitions and some $\gamma$-ray background lines. The spectrum is the sum of several runs irradiated at a proton beam energy ${\rm E_p}=205$~keV with a total charge of 75~C.}\label{190-keV}
\end{figure*}

In a previous publication \cite{Limata10_PRC} we have reported on a high precision measurement of the ${\rm E}= 214$, 304, and 326 keV resonances in the reactions $^{24}$Mg(p,$\gamma$)$^{25}$Al, $^{25}$Mg(p,$\gamma$)$^{26}$Al, and $^{26}$Mg(p,$\gamma$)$^{27}$Al, respectively. These studies were performed using multiple experimental approaches, i.e. $\gamma$-ray spectroscopy as well as accelerator mass spectrometry (AMS). The present work is the continuation of these efforts towards the important low-energy resonances in $^{25}$Mg(p,$\gamma$)$^{26}$Al.

\section{Experimental setup}
\label{expsetup}

The experiment has been performed at the 400~kV LUNA (Laboratory for Underground Nuclear Astrophysics) accelerator in the Laboratori Nazionali del Gran Sasso (LNGS) underground laboratory in Italy \cite{review,Broggini10_ARNPS}. The 1400 m rock overburden (corresponding to 3800 meter water equivalent) of the underground laboratory reduces the $\gamma$-ray background by more than three orders of magnitude for energies higher than 3.5 MeV, compared with a measurement on Earth's surface \cite{Bemmerer05}. The absolute $^{25}$Mg(p,$\gamma$)$^{26}$Al resonance strengths of the 92 and 189~keV resonances as well as an upper limit for the 130~keV resonance were measured with a 4$\pi$ BGO summing crystal. In addition, the 189~keV resonance was also studied with a HPGe detector allowing for a precise determination of the corresponding resonant branching ratios.

The details of the experimental setup are as reported previously \cite{Limata10_PRC}. Briefly, the 400~kV LUNA accelerator \cite{Formicola03} provided a proton current on target of up to 250 $\mu$A. The proton beam passed through several focussing apertures and a copper shroud, connected to a cold trap cooled to liquid nitrogen temperature, extending to within 2 mm from the target. The target plane was oriented perpendicularly to the beam direction.
A voltage of $-300$ V was applied to the cold trap to minimize emission of secondary electrons from both the target and the last aperture; the precision in the current integration was estimated to be about 2\%. The beam profile on target was controlled by sweeping the beam in the x and y directions within the geometry of the apertures. The targets were directly water cooled in order to prevent any heat damage during the measurements.

The BGO detector has a coaxial central hole (hole diameter = 6~cm; crystal length = 28~cm; radial thickness = 7~cm) and is optically divided in six sectors, each covering a 60$^\circ$ azimuthal angle (for details on detector and data acquisition see \cite{Limata10_PRC,Casella02}). The detector was mounted on a movable carriage such that the target could be placed in the center of the borehole maximizing the efficiency of the setup. For the HPGe measurement the target holder was replaced by a tube that allowed for an orientation of the target with its normal at 55$^\circ$ with respect to the beam direction. The HPGe detector (115$\%$ relative efficiency) was placed with its front face parallel to the target at a distance between target and detector front face of $\rm d=3.5$ and 6.0~cm, respectively. The detector was surrounded by 5 cm of lead, which reduced the background in the low-energy range by a factor 10.

The $^{25}$Mg targets have been produced from enriched $^{25}$MgO (enrichment $99\pm1$~\%). The $^{25}$MgO was mixed with Ta powder, heated with an electron-beam and evaporated on Ta backings (thickness 0.3~mm). The nominal $^{25}$Mg target density, $\simeq65$~$\mu$g/cm$^2$, corresponded to a 30~keV thickness at $\rm E_p=100$~keV \cite{srim,SRIMNIMB}. The stoichiometry and stability of each enriched target was frequently monitored scanning the 304~keV  $^{25}$Mg(p,$\gamma$)$^{26}$Al resonance during the course of the experiment. Finally, from our previous work \cite{Limata10_PRC} and other direct measurements, e.g. \cite{IL90}, it was obvious that oxidation represents a major difficulty in experiments with pure Mg targets.
In Ref. \cite{Limata10_PRC} the effect of oxygen contamination ($\approx10$~\% O abundance) in case of natural Mg targets was extensively studied, e.g. by means of Rutherford back scattering (RBS) analyses. The enriched Mg targets used in the present experiment had similar characteristics and a typical O abundance of 15~\% was found.

\section{Analysis and results}
\label{analysisresults}

The absolute strength $\omega\gamma$ of a narrow resonance can be obtained from a thick-target yield \cite{Rolfs88}.
In the present analysis the $^{25}$Mg(p,$\gamma$)$^{26}$Al low-energy resonance strengths were related to the recommended value of the 304~keV resonance, $\omega\gamma_{304}=30.8\pm1.3$~meV \cite{Limata10_PRC}, in $^{25}$Mg(p,$\gamma$)$^{26}$Al. The 304~keV resonance scans were performed before and after a long run on a low-energy resonance, at least after an accumulated charge of 10~C, under the same experimental conditions. Consequently from the thick-target yield a low-energy resonance strength $\omega\gamma_{\rm R}$ is obtained through the relation:
\begin{equation}\label{lowenergystrength}
\omega\gamma_R=3.29\times10^{-3}{\rm E_R}\frac{\rm Y_R}{\rm Y_{304}}\frac{\rm \varepsilon_{Mg}(E_{R,lab})}{\rm \varepsilon_{Mg}(317~keV)}z~\omega\gamma_{304}
\end{equation}
where E$_{\rm R}$ is the resonance energy in keV, $\varepsilon_{\rm Mg}({\rm E})$ the stopping power of protons in magnesium at the particular resonance energy evaluated in the laboratory frame, and Y$_{\rm R}$ and Y$_{304}$ the efficiency and dead time corrected yields of the studied resonance and the normalization to the 304~keV resonance, respectively. The correction factor {\it z} takes into account the exact energy dependence of the effective stopping power ratio between the two resonances due to the oxygen contamination. This correction is small, $z=0.9974$ (1.0044) for the 92~keV (189~keV) resonance, and was calculated for 15\% O abundance \cite{SRIMNIMB}. Its variation is less than $\pm0.3$\% over a large range of O content. A stoichiometry change due to the target bombardment was considered by averaging the yields of the normalization runs before and after the low-energy run. The data were used only in case this stoichiometry change was smaller than 10~\% (20~\% for the 92~keV resonance). Additionally, the beam energy for each on-resonance measurement was selected based on the 304~keV resonance scans ensuring that the resonance was placed well within the target layer.

The measurement of the 189~keV resonance with the HPGe detector was performed in a relatively close geometry. Therefore, a significant summing effect is expected from the large solid angle covered by the detector. The detection efficiency and the influence of summing effects have been studied in detail as described in Ref. \cite{Limata10_PRC}. In particular, all $\gamma$-ray lines of each $^{26}$Al level populated by the 189~keV resonance were observed and peak intensities analyzed. Subsequently, the summing-in and summing-out corrections have been calculated concurrently with the resonance strength in a simultaneous fit using the approach of Ref. \cite{imbriani05} including the data obtained at two distances. In this approach the resonance strength is a free parameter and the primary branching ratios (Table \ref{190-branchings}) as well as the summing correction are determined. The summing correction was less than 2~\% for most of the transitions and at maximum 13~\% (8.5~\%) for the ground state transition at $\rm d=3.5$~cm (6~cm).  Figure \ref{190-keV} shows the $\gamma$-ray spectrum obtained on top of the 189~keV resonance.

The beam induced background in the present experiment, e.g. from $^{11}$B and $^{19}$F contaminations in the target, was much smaller than found in previous works, e.g. Ref. \cite{IL90} (sample spectrum given in \cite{Iliadis}). This advantage, together with the larger volume of the HPGe detector and the low environmental background, resulted in a higher sensitivity of the branching ratio determination than previously available.
The deduced branching ratios are largely in agreement with previous works, although some significant differences were observed, e.g. the transition R$\rightarrow4705$~keV compared to Ref. \cite{IL90}. The transitions to the state at 5883~keV could not be verified while the 5726~keV transition (both discovered in Ref. \cite{IL90}) was observed with a reduced intensity. The primaries of these transitions overlap with strong natural $\gamma$-ray background lines from $^{214}$Bi (Radium series) at E$_\gamma=609.3$ and 768.4~keV, respectively,
and probably have been misinterpreted in Ref. \cite{IL90}. In contrast to previous measurements an additional ground state transition could be clearly identified after summing-in correction. Note the ground state transition, E$_\gamma=6496$~keV, is hampered by low detection efficiency at higher $\gamma$-ray energy and, therefore, could not be observed previously due to the cosmic-ray induced background.
Secondary transitions which could be analyzed are usually in good agreement (within 5\%) with Ref. \cite{EN88} except for the E$_{\rm x}=2070$~keV state where a significantly lower intensity for the transition to the 1058~keV state was observed ($2070\rightarrow0$: 2.3~\%, $\rightarrow417$: 35.1~\%, $\rightarrow1058$: 62.5~\%).

\begin{table}
\caption{Primary $\gamma$-ray branching ratios (in \%) of the ${\rm E}=189$~keV $^{25}$Mg(p,$\gamma$)$^{26}$Al resonance from the present HPGe measurement compared to a previous work. Energies are given in keV.}\label{190-branchings}
\begin{minipage}{\columnwidth}
\renewcommand{\footnoterule}{}
\renewcommand{\thefootnote}{\alph{footnote}}
\newcommand\T{\rule{0pt}{3.1ex}}
\newcommand\B{\rule[-1.7ex]{0pt}{0pt}}
\begin{tabular}{lr@{ $\pm$ }lr@{ $\pm$ }lr@{ $\pm$ }lr@{ $\pm$ }l}
\hline
\rule{0pt}{2.5ex}
E$_{\rm x}$  & \multicolumn{2}{c}{present work} & \multicolumn{2}{c}{\cite{IL90}} & \multicolumn{2}{c}{\cite{EN88}} & \multicolumn{2}{c}{\cite{Elix79_ZPA}} \T \B \\
\hline
5883  & \multicolumn{2}{c}{}     & 8  & 2                & \multicolumn{2}{c}{} & \multicolumn{2}{c}{} \T  \\
5726  & 3.0  & 1.0    & 7  & 2                & \multicolumn{2}{c}{} & \multicolumn{2}{c}{}     \\
4705  & 48.7 & 2.0                 & 35 & 4                & 50  & 2   & 50 & 8 \\
3403  & 1.5  & 0.5               & \multicolumn{2}{c}{$<3\footnotemark[1]$} & 4.5 & 0.9 & \multicolumn{2}{c}{} \\
3074\footnotemark[2] & 1.4 & 0.5 & \multicolumn{2}{c}{$<4$} & \multicolumn{2}{c}{} & 16 & 7 \\
2545  & 7.7  & 1.0               & 12  & 4               & 5.8 & 1.2 & \multicolumn{2}{c}{} \\
2365  & 21.9 & 2.0               & 26  & 4               & 19  & 1   & 17 & 7 \\
417   & 10.2 & 1.0               & 12  & 4               & 21  & 3   & 17 & 9 \\
0     & 5.6  & 1.1               & \multicolumn{2}{c}{}  & \multicolumn{2}{c}{} & \multicolumn{2}{c}{} \B  \\
\hline
\end{tabular}
\footnotetext[1]{Upper limit from Ref. \cite{Iliadis}.}
\footnotetext[2]{Yield observed is partly explained with the double-escape peak from $^{11}$B(p,$\gamma)^{12}$C (see \cite{IL90}). In the present spectra both lines are well separated and could be analyzed.}
\end{minipage}
\end{table}

The resulting total resonance strength, $\omega\gamma_{189}^{\rm  HPGe}= (9.0 \pm 0.4)\times10^{-7}$~eV, includes a probability for forming the $^{26}$Al ground state of $f_0=75\pm2$~\%. The ground state branching ratio was obtained from the full decay scheme of this resonance including all known secondary transitions. This value is consistent with an older literature ratio, $f_{\rm Endt}=75\pm1$~\% \cite{Endt87_NPA} (used in Ref. \cite{angulo}), while it disagrees with a more recent one, $f_{\rm Iliadis}=66$~\% \cite{IL96} (used in Ref. \cite{Iliadis10_NPA2}). However, the agreement with the former is only accidental due to the simultaneous revision of the primary branchings (Table \ref{190-branchings}) and the 2070~keV state branchings (see above). This indicates that the uncertainty on the ground state fraction in Ref. \cite{Endt87_NPA} is underestimated. The uncertainty of the $\omega\gamma$ determination includes relative detection efficiency (1.5~\%), statistical uncertainty (3~\%), and the uncertainty related to the normalization runs, e.g. stoichiometry change (3~\%). Uncertainty contributions common to both approaches, i.e. HPGe and BGO, were added after averaging and will be discussed below.

The $\gamma$-ray spectra from the BGO detector measurement were analyzed by means of a Monte Carlo simulation based on GEANT4 \cite{geant}. The Monte Carlo code uses as input parameters the available information on the $\gamma$-ray branchings of all transitions to the ground state or the isomeric first excited state of $^{26}$Al, e.g. \cite{EN88},  and tracks the simulated $\gamma$-ray cascades through setup and detector. The results are a sum spectrum of the total detector and a sum of the 6 single crystal segments, called single sum spectrum. The reliability and precision of this method has been successfully employed for $\omega\gamma$ determinations in (p,$\gamma$) reactions, e.g. the 304~keV resonance in $^{25}$Mg(p,$\gamma$)$^{26}$Al \cite{Limata10_PRC}.

\begin{figure}
\includegraphics[angle=0,width=\columnwidth]{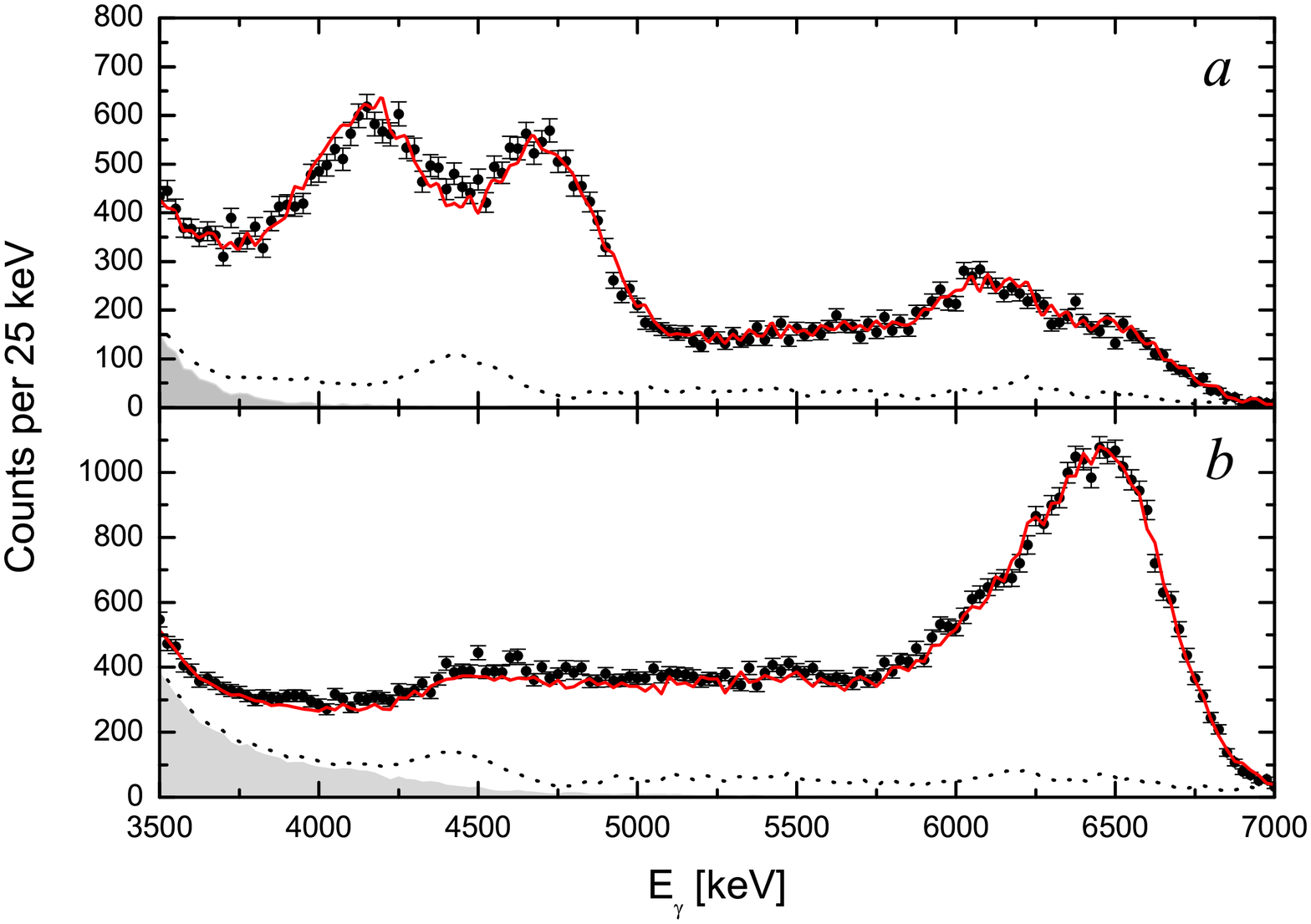}
\caption{The BGO $\gamma$-ray spectra (filled circles) on the ${\rm E}= 189$~keV $^{25}$Mg(p,$\gamma$)$^{26}$Al resonance. $a$) single sum (of all 6 crystals); $b$) total sum spectrum. The shaded area, dotted line, and solid red line represent the environmental background, the total background, and the complete yield fit including background and GEANT4 simulation, respectively. The spectrum was obtained after a target irradiation with $\approx10$~C at ${\rm E_p}=205$~keV. (For interpretation of the references to color in this figure legend, the reader is referred to the web version of this Letter.)}\label{BGO_190}
\end{figure}

The 189~keV resonance is located between strong resonances of $^{11}$B(p,$\gamma)^{12}$C, $^{18}$O(p,$\gamma)^{19}$F, and $^{19}$F(p,$\alpha\gamma)^{16}$O, respectively. As a consequence, the beam-induced background is not negligible in contrast to the previously studied resonances \cite{Limata10_PRC}. The reactions $^{11}$B(p,$\gamma)^{12}$C and $^{18}$O(p,$\gamma)^{19}$F have Q-values larger than $^{25}$Mg(p,$\gamma$)$^{26}$Al. Therefore, an energy cut restricting the accepted events to those with a total energy release in the detector of E$_{\gamma}<7$~MeV reduces strongly the background in the single sum spectrum. The remaining background contribution can be estimated with a simulation fitted to the high energy part of the spectrum, where the same energy cut was applied in the simulation of the background reactions. The background contribution of $^{19}$F(p,$\alpha\gamma)^{16}$O with a $\gamma$-ray line at E$_{\gamma}=6.13$~MeV can be determined from the shape of the experimental spectra only. However, the shape puts a strong constraint on the intensity of this line (less than 3~\% of the total yield).

The sum of the $^{25}$Mg(p,$\gamma$)$^{26}$Al simulation, the beam-induced background simulations and the environmental background, scaled from a long measurement without beam, were fitted to the experimental spectra, i.e. single and total sum. In both cases a good agreement was achieved (Fig. \ref{BGO_190}). This analysis of the BGO measurement led to a 189~keV resonance strength of $\omega\gamma_{189}^{\rm BGO}= (9.0 \pm 0.5)\times10^{-7}$~eV including uncertainties of 3~\% for each simulation, i.e. 189 and 304~keV resonances as well as background, and the stoichiometry variation (2~\%) while the statistical uncertainty is well below 0.5~\%. The results of both detection approaches are in excellent agreement and result in a weighted average of $\omega\gamma_{189}= (9.0 \pm 0.6)\times10^{-7}$~eV including also common uncertainties, e.g. the normalization uncertainty, $\omega\gamma_{304}$ (4.2~\%), energy dependence of the stopping power (1.5~\%), and current measurement (2~\%).
This value is higher than the result of a previous work, $\omega\gamma= (7.4 \pm 1.0)\times10^{-7}$~eV \cite{IL90}\footnote{Note that in NACRE \cite{angulo} a value of $\omega\gamma= (7.1 \pm 1.0)\times10^{-7}$~eV is quoted for Ref. \cite{IL90}. A more recent compilation \cite{Iliadis10_NPA2} uses $\omega\gamma= (7.2 \pm 1.0)\times10^{-7}$~eV as a normalization to reference strengths from Ref. \cite{Iliadis01_ApJS}.}. However, the difference is not surprising since there is a discrepancy in the most intense primary transition for this resonance (Table \ref{190-branchings}). Possible explanations might be the influence of an angular distribution or an incomplete summing-out correction in the 0$^\circ$ measurement of \cite{IL90}. 
The fact that the present HPGe experiment delivered consistent results at two different distances indicates that the summing corrections are well under control and angular distribution effects are negligible at 55$^\circ$. Furthermore, the good agreement of the simulated $\gamma$-ray spectrum - based on the branching ratios from the HPGe phase - with the observed BGO spectrum supports this assumption as well. As a consequence we discard the result of Iliadis et al. \cite{IL90} and recommend the present value. Finally, the present value is incompatible with a resonance strength, $\omega\gamma_{\rm AMS}=(1.5\pm0.2)\times10^{-7}$~eV (corrected for total strength), measured by means of AMS \cite{arazi}. This value can be excluded for reasons that have been discussed previously \cite{Limata10_PRC}.

\begin{figure}
\includegraphics[angle=0,width=\columnwidth]{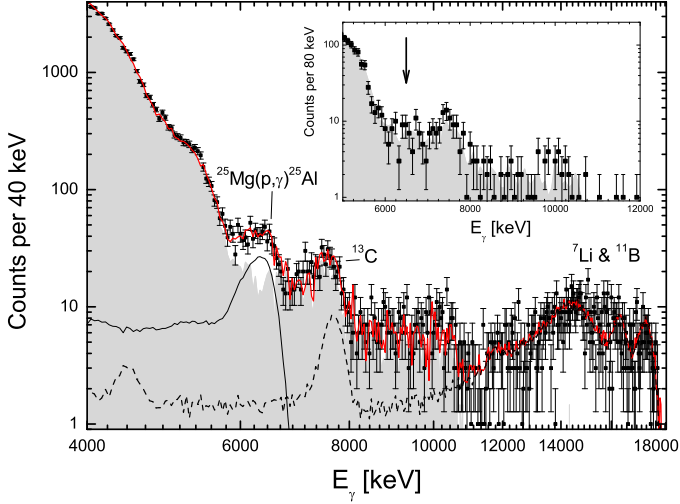}
\caption{The BGO $\gamma$-ray total sum spectrum on the 92~keV $^{25}$Mg(p,$\gamma$)$^{26}$Al resonance (${\rm E_p}=100$~keV). The shaded area, thin solid line, and solid red line represent the environmental background, the $^{25}$Mg(p,$\gamma$)$^{26}$Al GEANT4 simulation, and the total yield fit including background and simulation, respectively. The contributions from contaminant reactions (dashed line from GEANT4 simulations) are labeled. The insert shows the background at ${\rm E}=86.5$~keV, i.e. below the resonance. The region of the sum peak is indicated by an arrow. (For interpretation of the references to color in this figure legend, the reader is referred to the web version of this Letter.)}\label{BGO_90}
\end{figure}

\begin{figure}
\includegraphics[angle=0,width=\columnwidth]{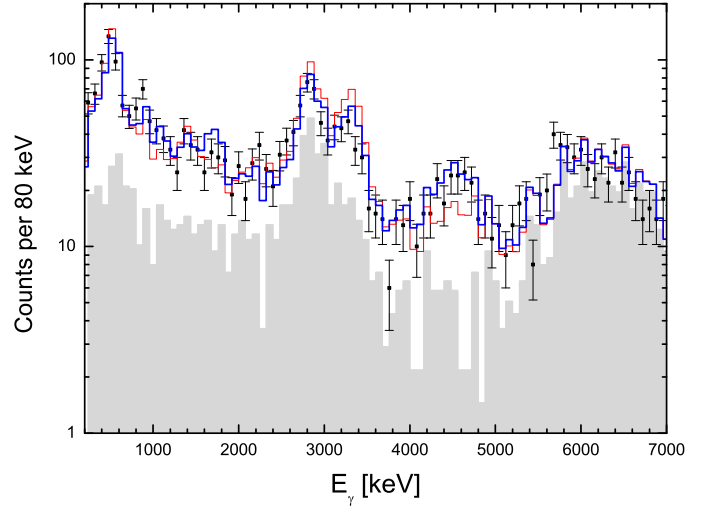}
\caption{The BGO single sum spectrum on the 92~keV resonance. The shaded area represents the environmental background, while the thin red (thick blue) line shows the total yield fit including background and simulation using the primary branching from Ref. \cite{CH83a} (alternative branching, see text). The step at $\rm E_\gamma=5.6$~MeV is caused by the analysis cut. (For interpretation of the references to color in this figure legend, the reader is referred to the web version of this Letter.)}\label{BGO_90single}
\end{figure}

In spite of the large sum peak efficiency of the BGO detector and the high beam intensity only the 92~keV resonance could be clearly observed below $\rm E=189$~keV. In addition, for the 130~keV resonance an upper limit was obtained while all other low-energy resonances are too weak to be studied with the present experimental setup. The resonance strength of the 92~keV resonance was determined from count rate and peak shape in the energy region E$_\gamma=5.6 - 7.0$~MeV of the total sum spectrum (Fig. \ref{BGO_90}). This region was also used to gate the single sum spectrum: only reconstructed events depositing their total energy in this gate window were accepted in the single spectrum (Fig. \ref{BGO_90single}). This procedure strongly reduced the background and provided indications on the resonant $\gamma$-ray branching ratios. Although the obtained information were not conclusive due to the limited statistics, a simulation based on available literature data, i.e. the primary branchings of Ref. \cite{CH83a}, showed significant differences to the experimental spectrum (Fig. \ref{BGO_90single}, thin solid line). Simulations varying the primary branchings suggested that the 92~keV resonance has stronger transitions through the 1851 and 2070~keV states in $^{26}$Al (Fig. \ref{BGO_90single}, thick solid line). However, these changes influence the ground state fraction only (see below). Systematic studies of the total sum peak efficiency as well as the peak shape revealed only a small effect of the exact branching ratio on these quantities \cite{Best}.

The sources of beam induced background remain the same as discussed above down to $\rm E\approx 110$~keV. At lower energies the environmental background (see Ref. \cite{Bemmerer05} for details) is the dominating background component and was found to be constant within the statistical uncertainty over the entire period of the experiment. Minor additional background contributions arise from $^7$Li(p,$\gamma)^8$Be and $^{13}$C(p,$\gamma)^{14}$N (see Fig. \ref{BGO_90}). Below the 92~keV resonance the observed spectrum is basically identical to the environmental background (see insert Fig. \ref{BGO_90}). This is a strong indication that the beam induced background subtraction is robust and no unexpected component is hidden below the sum peak. The corresponding spectra of the identified background components have been simulated with the GEANT4 code and were fitted together with the simulated $^{25}$Mg(p,$\gamma)^{26}$Al resonant spectrum and the time normalized environmental background.

The fit resulted in 410 events from the $^{25}$Mg(p,$\gamma)^{26}$Al reaction in the region of interest. For the total run time of 24 days and the collected charge of $\approx370$~C this is equivalent to 2 reactions per hour. The corresponding strength of the 92~keV resonance is $\omega\gamma_{92}=(2.9\pm0.6)\times10^{-10}$~eV with a sum peak efficiency of 38~\%. This result is to our knowledge the lowest ever directly measured resonance strength. The quoted uncertainty includes the statistical uncertainty (9~\%, which takes into account the statistical uncertainty of the background) as well as uncertainties due to beam induced background, i.e. equal to the number of subtracted counts (5~\%), the choice of the fit region (5~\%), and the GEANT4 simulation with a not precisely known branching ratio (5~\%). Additional contributions arise from the correction due to the stoichiometry variation including charge integration (15~\%), the normalization to $\omega\gamma_{304}$ (4.2~\%), and the relative stopping power data (1.5~\%). The uncertainty in the resonance energy has been neglected.

Finally, at such low energies the electron screening effect \cite{Assenbaum87_ZPA} has to be considered. The resonance strength is proportional to the penetrability P$_l({\rm E})$ of the orbital angular momentum $l$ through the proton partial width $\Gamma_{\rm p}$, $\omega\gamma\propto\Gamma_{\rm p}\propto{\rm P}_l({\rm E})$. Thus, the enhancement factor f$_{\rm es}$ of the entrance channel can be expressed as:
\begin{equation}\label{screening}
{\rm f_{es}}=\frac{\omega\gamma_{\rm screen}}{\omega\gamma_{\rm bare}}=\frac{{\rm P}_l({\rm E+U_e})}{{\rm P}_l({\rm E})}
\end{equation}
where for small $l$ the usual approximation $\rm f_{es}\approx \exp(\pi\eta{\rm U_e/E})$ is still valid \cite{Assenbaum87_ZPA}. The screening potential U$_{\rm e}=1.14$~keV was calculated in the adiabatic limit from atomic binding energies \cite{Huang76_ADNDT} leading to an enhancement factor f$_{\rm es}=1.25$ for the 92~keV resonance (Table \ref{omegagamma_comparison}).

The present result corrected for screening is within the quoted uncertainty in agreement with the NACRE value, $\omega\gamma_ {92}^{\rm NACRE}=(1.16^{+1.16}_{-0.39})\times10^{-10}$~eV \cite{angulo}. The NACRE value is based on the average proton partial width from a reanalysis \cite{IL96} of older proton stripping data \cite{Betts78_NPA,CH89,Rollefson90_NPA}. This proton width, $\Gamma_{\rm p}=(2.8\pm1.1)\times10^{-10}$~eV (uncertainty as quoted in \cite{Iliadis10_NPA3}), is also used in Ref. \cite{Iliadis10_NPA2}. The proton width from the present experiment, $\Gamma_{\rm p}=(5.6\pm1.1)\times10^{-10}$~eV, deviates from this literature value by 1.8$\sigma$, where $\sigma$ is calculated as the quadratic sum of the individual uncertainties. Therefore, at the 90~\% confidence level the literature value is incompatible with the present result, while the present proton width is in good agreement with the original value of Ref. \cite{Rollefson90_NPA}, $\Gamma_{\rm p}=(5.2\pm1.3)\times10^{-10}$~eV.

The ground state branching fraction is an additional important input parameter for the reaction rate. The available literature information in case of the 92~keV resonance are contradictory. Based on the experimental branching ratio determination through the $^{24}$Mg($^3$He,p$\gamma)^{26}$Al reaction \cite{CH83a} a probability of $80\pm20$~\% was deduced \cite{CH83b}. The same authors quote in Ref. \cite{CH86} a value of 61~\% while the compilation of Endt and Rolfs \cite{EN87} gives 85~\%. The additional information from the single sum spectrum suggested stronger transitions through $^{26}$Al states that predominately decay to the isomeric state reducing the ground state fraction. The ground state probability in the present work is calculated from the full decay scheme including the new information and a value of $60^{+20}_{-10}$~\% is recommended.

The measurement on the 130~keV resonance (total charge $\approx90$~C at $\rm E_p=140$~keV) was analyzed in the same way, but due to the stronger beam induced background in this proton energy region the signal could not be clearly identified. As a consequence only an upper limit, $\omega\gamma_{130}<2.5\times10^{-10}$~eV, is quoted here. This result is in agreement with the value given in NACRE, $\omega\gamma_{130}^{\rm NACRE}<1.4\times10^{-10}$~eV \cite{angulo} and verifies the information from transfer reactions that this resonance is of no astrophysical relevance.

\begin{table}
\caption{The new recommended $^{25}$Mg(p,$\gamma$)$^{26}$Al resonance strengths (uncorrected for screening) and corresponding ground state fractions $f_0$. The electron screening enhancement factor f$_{\rm es}$ was calculated according to \cite{Assenbaum87_ZPA}.}\label{omegagamma_comparison}
\begin{minipage}{\columnwidth}
\renewcommand{\footnoterule}{}
\renewcommand{\thefootnote}{\alph{footnote}}
\newcommand\T{\rule{0pt}{3.1ex}}
\newcommand\B{\rule[-1.7ex]{0pt}{0pt}}
\begin{tabular}{c c c c}
\hline
 E [keV]\footnotemark[1] & $\omega\gamma$ [eV] & f$_{\rm es}$ & $f_0$ \T \B \\
\hline
  304.0 & $(3.08\pm0.13)\times10^{-2}$ \footnotemark[2] & 1.04 & $0.878\pm0.010$\footnotemark[2] \T \\
  189.5 & $(9.0\pm0.6)\times10^{-7}$    & 1.08 & $0.75\pm0.02$   \\
  130.0 & $<2.5\times10^{-10}$           & 1.14 & $0.73\pm0.01$\footnotemark[3]  \\
  92.2  & $(2.9\pm0.6)\times10^{-10}$  & 1.25 & $0.6^{+0.2}_{-0.1}$ \B  \\
\hline
\end{tabular}
\footnotetext[1]{from Ref. \cite{EN87}, the uncertainty is less than 0.2~keV in all cases}
\footnotetext[2]{from Ref. \cite{Limata10_PRC}}
\footnotetext[3]{from Ref. \cite{EN87} (Note the uncertainty might be underestimated)}
\end{minipage}
\end{table}

\section{Summary and conclusion}

The resonance strengths of the 92 and 189~keV resonances in $^{25}$Mg(p,$\gamma$)$^{26}$Al have been measured with unprecedented sensitivity relative to the well-known 304~keV resonance which has been established in the previous study \cite{Limata10_PRC}. In particular, the first direct study of the 92~keV resonance largely reduces the uncertainty in the reaction rate. The results are summarized in Table \ref{omegagamma_comparison}. The experimental $\omega\gamma$ values are given together with the ground state branching fraction and the electron screening correction. Note that due to the relatively large Z of the target nucleus the influence is already sizeable for the 304~keV resonance. The nuclear reaction rate for $^{25}$Mg(p,$\gamma$)$^{26}$Al and the implication on astrophysical models of the $^{26}$Al synthesis will be discussed in a forthcoming publication.

The authors are grateful to H.~Baumeister (Institut f\"ur Kernhysik, Universit\"at M\"unster, Germany) and M. Loriggiola (INFN Legnaro, Padova, Italy) for the excellent preparation of $^{25}$Mg targets. The present work has been supported by INFN and in part by the EU (ILIAS-TA RII3-CT-2004-506222), OTKA (K101328 and K68801), and DFG (Ro~429/41).

\bibliographystyle{model1a-num-names}
\biboptions{compress}
\bibliography{letter} 

\end{document}